\documentclass[aps,prb,showpacs,twocolumn,superscriptaddress]{revtex4-1}
\usepackage{graphicx,amssymb,amsmath,bm,color,flushend}

\newcommand{\jav}[1]{{#1}}

\begin{document}
\definecolor{darkgreen}{rgb}{0,0.5,0}
\definecolor{matlabmagenta}{rgb}{1,0,1}

\title{Geometrical quench and dynamical quantum phase transition in the $\alpha-T_3$ lattice}

\author{Bal\'azs Gul\'acsi}
\email{gulacsi@phy.bme.hu}
\affiliation{Department of Theoretical Physics and MTA-BME Lend\"ulet Topology and Correlation Research Group,Budapest University of Technology and Economics, 1521 Budapest, Hungary}
\author{Markus Heyl}
\affiliation{Max-Plank-Institute f\"ur Physik Komplexer Systeme, N\"othnitzer Strasse 38, 01187 Dresden, Germany\looseness=-1}
\author{Bal\'azs D\'ora}
\affiliation{Department of Theoretical Physics and MTA-BME Lend\"ulet Topology and Correlation Research Group,Budapest University of Technology and Economics, 1521 Budapest, Hungary}
\date{\today}

\begin{abstract}
We investigate  quantum quenches and the Loschmidt echo in the two dimensional,  three band $\alpha-T_3$ model, a close descendant of the dice lattice. 
By adding a chemical potential to the central site, the integral of the Berry curvature of the bands in different valleys is continously tunable by the ratio of the hopping integrals between the sublattices.
By investigating one and two filled bands, we find that  dynamical quantum phase transition (DQPT), i.e. nonanalytical temporal behaviour in the rate function of the return amplitude, occurs
for a certain range of parameters, independent of the band filling.
By focusing on the effective low energy description of the model, we find that DQPTs happen not only in the time derivative of the rate function, which is a common feature in two dimensional models, but in the rate function itself.
This feature is not related to the change of topological properties of the system during the quench, but rather  follows from population inversion for all momenta.
This is accompanied by the appearance of dynamical vortices in the time-momentum space of the Pancharatnam geometric phase. The positions of the vortices form an infinite vortex ladder, i.e. a macroscopic phase structure, which allows us to identify the dynamical phases that are separated by the DQPT.
\end{abstract}

\maketitle

\section{Introduction}
The swift progress of measurement techniques, especially ultra-cold atoms in optical lattices\cite{Bloch2012,Geiger2015}, have opened the way for the experimental study of real-time dynamics of closed quantum many-body systems. 
This includes the observation of exotic phenomena such as prethermalization\cite{Schmiedmayer2012}, many body localization\cite{Bloch2015}, particle-antiparticle production in gauge theories\cite{Martinez2016}.
Motivated by these developments a great amount of theoretical works concerning nonequilibrium phenomena have been conducted in the recent years. One of the more popular method for driving a system out of equilibrium is a 
sudden quantum quench. In a quantum quench protocol a closed quantum system is prepared in an eigenstate $|\Psi_0\rangle$ of some initial Hamiltonian $H_0$ and then have the system dynamically evolve in time under a different 
Hamiltonian $H$. An approach to understand dynamics after a quantum quench is to study the behaviour of the return amplitude
\begin{gather}
G(t)=\langle \Psi_0|\Psi(t)\rangle=\langle \Psi_0|e^{-iHt}|\Psi_0\rangle.
\label{loch}
\end{gather}
There is a formal similarity between Eq.~\eqref{loch} and the partition function $Z=\text{Tr}(e^{-\beta H})$, where $\beta$ is the inverse temperature, which was first pointed out in Ref.~\cite{Heyl2013}. Based on this formal 
similarity the theory of dynamical quantum phase transitions (DQPT) have emerged\cite{Heyl2018}.

By statistical mechanics the information of a system's thermodynamical properties is contained within the partition function $Z$. The free energy density of the system is calculated from the partition function as 
$f=F/N=-(\beta N)^{-1}\ln Z$ with $N$ being the degrees of freedom of the system. A phase transition occurs when $f$ behaves nonanalytically as a function of a control parameter, such as the temperature or external magnetic field. 
Obviously, in a non-equilibrium situation, the partition function cannot be formulated in the conventional sense. 
The theory of DQPTs shifts the formal role of the partition function to the complex Loschmidt amplitude $G(t)$. A dynamical quantum phase transition is then defined as the nonanalytic behaviour as a 
function of time in the dynamical counterpart of the free energy density: $g(t)=-\lim_{N\to\infty}N^{-1}\ln G(t)$. Thus, mathematically a DQPT occurs at critical times ($t^*$), whenever $G(t^*)=0$. 
Physically, the return amplitude is related to the work done during the quench\cite{Silva2008}, which in principle makes DQPTs a measurable phenomenon. Recently, dynamical quantum phase transitions 
were observed in various experimental platforms including quantum simulators\cite{Jurcevic2017,Zhang2017,Exp2018,Exp2019}, nanomechanical systems\cite{Jiangfeng2019} and single photon resonators\cite{Xue2019}  .

The underlying theme of identifying DQPTs is to investigate under when the condition $G(t)=0$ holds.
 To this end, the zeros of the return amplitude, refered to as Fisher zeros, are studied in the complex 
plane: $G(z_n)=\langle \Psi_0|e^{-Hz_n}|\Psi_0\rangle=0$, where $z_n\in \mathbb C$.  In the thermodynamic 
limit the Fisher zeros coalesce into lines in one dimension or areas in higher spatial dimensions. The final task is to identify conditions for the Fisher zeros to cross the imaginary axis. 
DQPTs have been found in both integrable\cite{Heyl2013} and nonintegrable\cite{Karrasch2013} spin systems for quenches across quantum critical points. However, there are also reports 
of DQPTs that occur when a parameter is not quenched accross a quantum critical point\cite{Vajna2014,Schmitt2015}, denying the one-to-one correspondence to conventional phase transitions\cite{sachdev}. Furthermore, the appearance of DQPTs in topological band models are also being actively researched, however these are often limited to two band models\cite{Vajna2015}. 
For quantum quenches between gapped phases in a generic multiband system, a robust DQPT is a consequence of momentum-space zeros in the wave function overlap between the prequench state and all postquench 
energy eigenstates. Zeros in wave function overlaps are topologically protected if the topological indices of the two participating wave functions are different\cite{Huang2016}. 

In the present paper we investigate the occurance of DQPTs in the three band $\alpha-T_3$ lattice, depicted in Fig.~\ref{lattice}.
The main interesting features of this model are that one of the bands is always a zero energy flat band and the integral of the Berry curvature for the other bands of the gapped $\alpha-T_3$ lattice depend  continously on the ratio of the sublattice hopping amplitudes. 
These are detailed in Section~\ref{model}. Afterwards, we establish the properties of the quench scenarios and deduce the conditions for the DQPTs: \jav{population inversion for band states that are occupied in the prequench state.} In Section~\ref{gapless}., we detail the quench scenario, when a gap is created and the hopping ratio is changed, also the properties of the resulting DQPT is presented. Section~\ref{gap1}-\ref{gap2}. features the scenarios when the initial Hamiltonian is gapped.

\section{The model}\label{model} 
\begin{figure}
\includegraphics[width=4cm]{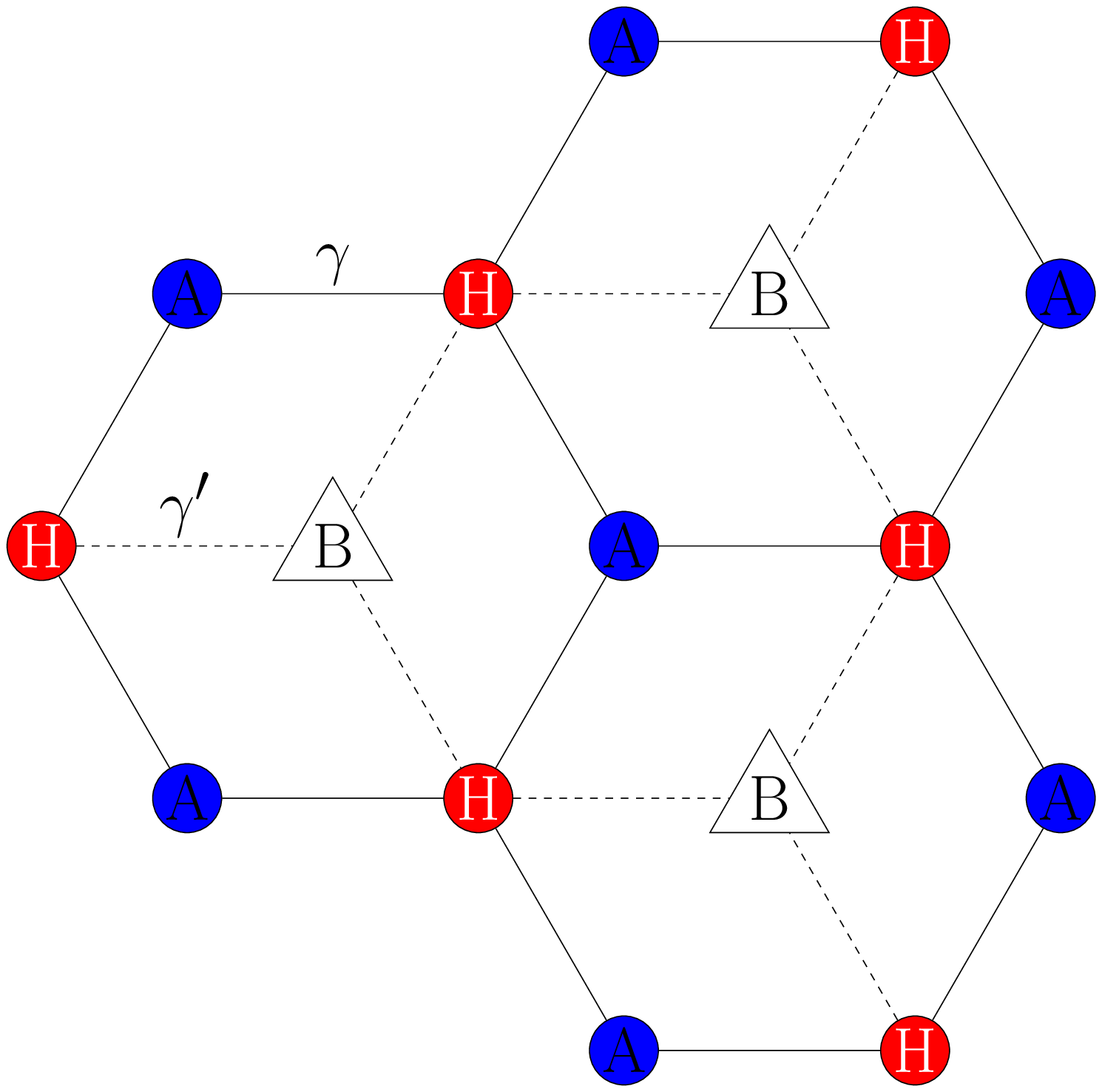}\hspace{0.6cm}\includegraphics[width=4cm,height=4cm]{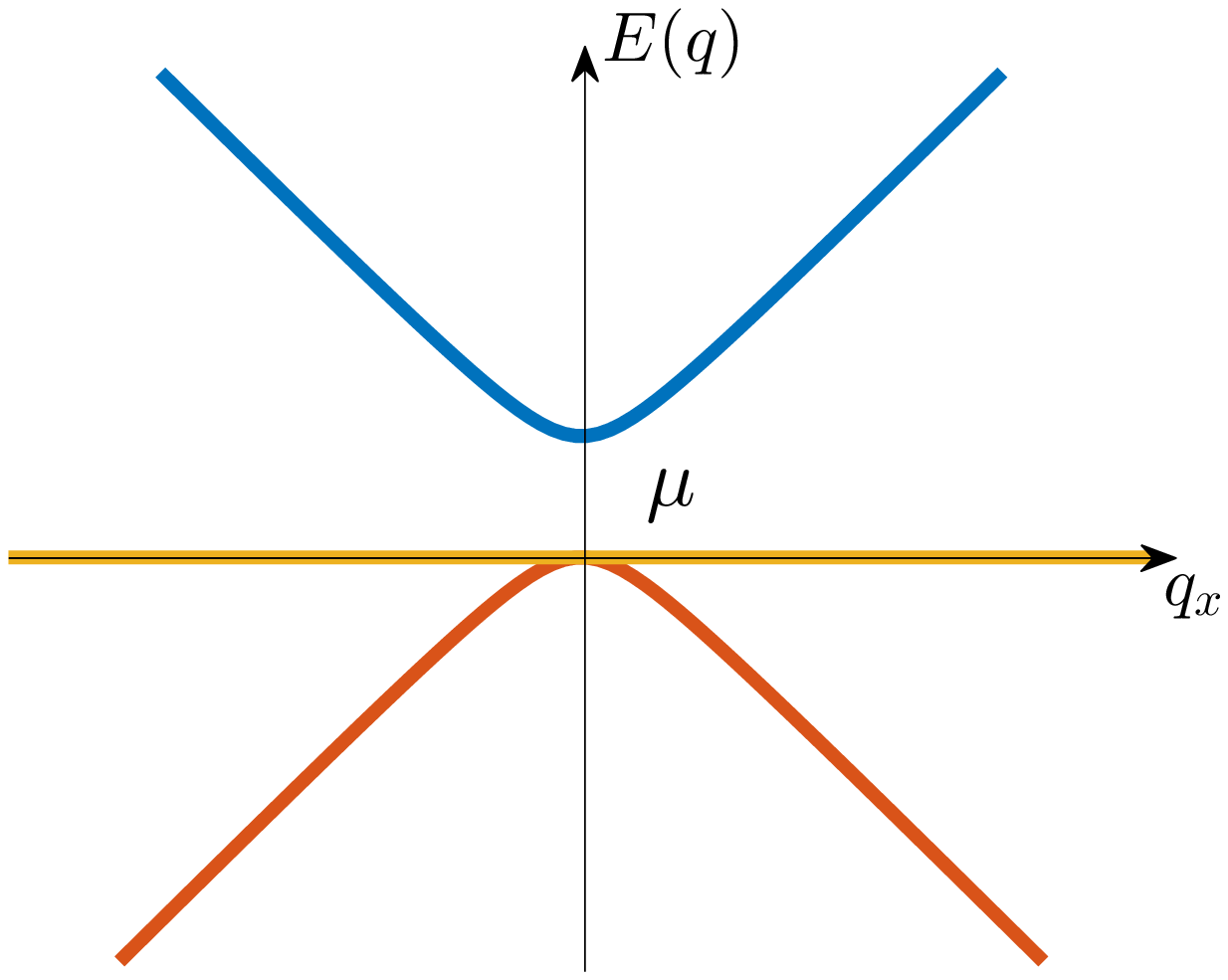}
\caption{Left panel: the $\alpha-T_3$ lattice can be considered as two interpenetrating the honeycomb lattices. There is no hopping between the two rim sites, labeled ($A,B$), however they are both connected to the hub site, labeled ($H$), 
with hopping amplitudes $\gamma$, $\gamma'$. 
Their ratio $\alpha=\gamma'/\gamma$ serves as a tuning parameter that distinguishes different lattice stuctures. When $\alpha=0$ we recover the familiar honeycomb lattice and with $\alpha=1$ the dice lattice. 
Right panel: the dispersion relation near a Dirac point with finite $\mu$ chemical potential on the hub site.}
\label{lattice}
\end{figure}

The $\alpha-T_3$ lattice is a two dimensional structure whose Bravais lattice is a triangular lattice and the unit cell contains three atoms. The basis includes two rim sites $(A,B)$ that are connected to a 
hub site $(H)$ with hopping amplitudes $\gamma$ and $\gamma'$, see Fig.~\ref{lattice}. 
Within the tight binding picture, the $\alpha-T_3$ lattice has three bands: a zero energy flat band and two other bands that touch each other at the corners of the hexagonal shaped first Brillouin zone. 
Similar to graphene\cite{grafrmp}, the energy bands are linear near these points and thus the system can be described around the Dirac points with a low-energy effective pseudospin-1 Hamiltonian\cite{Bercioux2009,latticegen,Raoux2014}.
By adding a local chemical potential to the central hub site\cite{dora2014,Cserti2017}, the Dirac points remain at the K points of the Brillouin zone\cite{grafrmp} 
and the low energy Hamiltonian reads as:
\begin{gather}
H_\xi(\mathbf q)=\begin{pmatrix}0 && v_Fq_\xi && 0 \\
v_Fq_\xi^* && \mu && \alpha v_F q_\xi\\
0&& \alpha v_F q_\xi^*&&0
\end{pmatrix}.
\label{ham}
\end{gather}
Here $v_F=3a\gamma/2$ is the Fermi velocity, with $a$ being the lattice constant, $\alpha=\gamma'/\gamma$ is the ratio of the hopping amplitudes. 
The momentum dependence is given in polar coordinates with $\xi=\pm1$ valley index: $q_\xi=q_x-i\xi q_y=q\exp(-i\xi\vartheta_q)$, 
where $q=\sqrt{q_x^2+q_y^2}$, $\vartheta_q=\arctan(q_y/q_x)$. Due to the addition of the local chemical potential $\mu$ to the hub site, the system becomes gapped and develops unique topology. For the Hamiltonian matrix in Eq.~\eqref{ham} the eigenvalues and eigenfunctions are:
\begin{gather}
E_{0}(\mathbf q)=0,\quad
E_{1,2}(\mathbf q)=\frac{\mu}{2}\pm\sqrt{\frac{\mu^2}{4}+v_F^2(1+\alpha^2)q^2},\nonumber\\
|\psi_0(\mathbf q)\rangle=\frac{1}{\sqrt{1+\alpha^2}}\begin{pmatrix}-\alpha e^{-i\xi\vartheta_q}\\0\\e^{i\xi\vartheta_q}\end{pmatrix},\nonumber\\ 
|\psi_{1,2}(\mathbf q)\rangle=\frac{1}{\sqrt{E_{1,2}(\mathbf q)^2+v_F^2(1+\alpha^2)q^2}}\begin{pmatrix} v_Fq_\xi \\ E_{1,2}(\mathbf q) \\ \alpha v_F q_\xi^*
\end{pmatrix}.
\label{eigen}
\end{gather}

For the complete characterization of the present model, the topological properties are presented through the integral of the Berry curvature for all bands:
\begin{gather}
C_j=\displaystyle\frac{i}{2\pi}\int d^2\mathbf q \nabla_\mathbf q\times\langle\psi_j(\mathbf q)|\nabla_\mathbf q\psi_j(\mathbf q)\rangle.\label{berry}
\end{gather}
Using Eq.~\eqref{eigen} the corresponding integrals in Eq.~\eqref{berry} are calculated as
\begin{gather}
C_0=0,\quad C_{1,2}=\pm\frac{\xi(\alpha^2-1)}{\alpha^2+1}(1-\delta_{\mu 0}).
\end{gather}
The topological number for the flat band is zero, while for the dispersive bands, $C_1+C_2=0$ always. However, in each valley, 
the integrals of the Berry curvature continously depend on the hopping ratio $\alpha$. This means that around a single valley the electrons occupy bands that can possess  non-integer  integrals of the Berry curvature, 
more precisely any real number in the $[-1,1]$ interval depending on the hopping ratio, although there is a finite gap in the system.
By considering both valleys, the non-integer integrals of the Berry curvature for the bands in each valley add up to integer Chern numbers for the bands.
As we will see, since the quench dynamics are independent from the valley index $\xi$, it is satisfactory to focus on a single valley. Henceforward, 
we are introducing the different quench scenarios in the $\alpha-T_3$ model and deduce conditions for the appearance of DQPTs.

\subsection{Quench protocols.} 
As the prequench state we are using the fully filled low energy band $|\psi_{in}(\mathbf q)\rangle=|\psi_2(\mathbf q)\rangle$ with starting parameters $\mu_0$ and $\alpha_0$. 
The post quench Hamiltonian that governs the time evolution has parameters $\mu$ and $\alpha$. 
The return amplitude is $G(t)=\prod_\mathbf{q} G(\mathbf q,t)$ with
\begin{gather}
G(\mathbf q,t)=\sum_{j=0,1,2}|\langle\psi_j(\mathbf q)|\psi_{in}(\mathbf q)\rangle|^2 e^{-iE_j(\mathbf q)t}.
\label{loch1}
\end{gather}
Hereafter, we use $p_j=|\langle\psi_j(\mathbf q)|\psi_{in}(\mathbf q)\rangle|^2$ for the overlaps between the prequench state and all of the postquench Hamiltonian eigenfunctions, which is the probability
of a given single particle state being occupied after the quench.
The return amplitude is a product with respect to the momentum $\mathbf q$, thus it is zero whenever $G(\mathbf q,t)=0$. 
Since $p_j\geq0$ for all $q$, Eq.~\eqref{loch1} can be interpreted as a sum of complex numbers with magnitude $p_j$ and phase $-E_jt$. 
The sum is zero whenever the complex numbers $p_je^{-iE_jt}$ form a closed polygon on the complex plane at times $t^*$. To form a polygon the overlaps must obey the triangle inequality, 
which combined with the fact that $\sum_j p_j=1$ results in a condition for DQPTs to occur, specifically:
\begin{gather}
\exists \mathbf q \neq 0: p_j(\mathbf q)\leq \frac{1}{2},\quad\forall j.
\label{condition}
\end{gather}
For the overlaps $\{p_j\}$ that satisfy Eq.~\eqref{condition} there exists solutions for the equation $\sum_j p_je^{-i\varphi_j}=0$ with some $\varphi_j$ phases. 
The only remaining question is to whether $E_jt$ can evolve into these $\varphi_j$ phases. The answer is affirmative, whenever the energy bands of the postquench Hamiltonian are rationally 
independent\cite{Huang2016,Budich2019,Note2}. 
In our case if $\mu\neq0$, i.e. the gap is not closed during the quench, this phase condition is automatically satisfied, and as such the only condition for 
the occurance of a DQPT is the time independent condition for the overlaps in Eq.~\eqref{condition}. The presence of the flat band is invalueable in satisfying Eq.~\eqref{condition}. As we will see a proportion of the electrons are frozen into the flat band after the quench and the dynamics play out only on the remaining part of the system. This greatly increases the chance for the band populations to satisfy Eq.~\eqref{condition}.

Due to rotational invariance of the low energy Hamiltonian in Eq.~\eqref{ham},
 the overlaps only depend on the magnitude of the momentum $|\mathbf q|=q$. The dynamical counterpart of the free energy, the rate function can be calculated, using Eq.~\eqref{loch1} as
\begin{gather}
g(t)=-\frac{1}{N}\sum_\mathbf q \ln G(\mathbf q,t)=-A_c\int \frac{\text d^2\mathbf q}{(2\pi)^2}\ln G(q,t)\implies\nonumber\\
g(t)=-\lim_{q_F\to\infty}\frac{A_c q_F^2}{2\pi q_F^2}\int_0^{q_F}\text{d}q\text{ } q \ln G(q,t),
\label{kisg}
\end{gather}where $A_c$ is the area of the unit cell and $q_F$ is a high energy cutoff in momentum space. We make the continuum limit with $\displaystyle\lim_{q_F\to\infty}A_c q_F^2/4\pi=1$, so that there is one electron per site. 


We argue that having two filled bands initially does not differ much from having a single filled band.
Preparing initially the ground state with some given chemical potential could mean that the flat band is also occupied. 
We argue that calculating with the two filled bands scenario would yield identical results to only one filled band. The two filled bands can be considered as starting from the the fully filled state 
(all bands are filled) and annihilating electrons from one band for all momenta to reach one empty band. 
Then, there is one empty hole band, which can be treated mathematically identically to having a single filled electron band. 
For our specific setting in the  $\alpha-T_3$, we have checked that all of our results, obtained for the single filled lowest energy band, remain intact for the case of two filled bands, i.e. when the flat band is also occupied, and only the high energy band is empty.

Since the system is translationally invariant, the momentum $\mathbf q$ is a good quantum number and the return amplitude factorizes to one particle contributions. 
If the state labeled by $\mathbf q$ is occupied and fulfills the condition in Eq.~\eqref{condition}, then it participates in the DQPT. 
This means that if the prequench state is a partially filled band that contains these states DQPTs happen, however the exact properties of these DQPTs are 
beyond the scope of the present investigation.

\section{Gapless to gapped quench scenario}\label{gapless}

First, we discuss the case when the prequench Hamiltonian has no gap $(\mu_0=0)$ and during the quench a gap opens $(\mu\neq0)$. As mentioned above, in this case the energies of the postquench bands are rationally independent, 
so we only need to analyze the overlap functions and their properties: 
\begin{gather}
p_0=\frac{(\alpha_0-\alpha)^2}{2(1+\alpha^2)(1+\alpha_0^2)},\nonumber\\
p_{1,2}=\frac{(-E_{1,2}\sqrt{1+\alpha_0^2}+v_F(1+\alpha\alpha_0)q)^2}{2(1+\alpha_0^2)(E_{1,2}^2+v_F^2(1+\alpha^2)q^2)}.
\label{m0=0}
\end{gather}
In Eq.~\eqref{m0=0}, the flat band overlap is a constant, furthermore $p_0<1/2$. The $p_0$ proportion of the electrons are frozen into the flat band after the quench. Since $p_1+p_2=1-p_0$ is also a constant, their derivatives with respect to $q$ have opposite 
signs: $\partial p_1/\partial q=-\partial p_2/\partial q$. This is important because we notice that $p_1(q=0)=1/2$ and $\partial p_1/\partial q<0$, $\forall q$, consequenlty $p_1<1/2$, if $q>0$. 
The determining factor is hence the overlap between the  low energy bands of the pre- and postquench Hamiltonians. For $p_2$ is an increasing function of $q$, and  to observe a DQPT we need to satisfy the condition
\begin{gather}
p_2(q\to0)=\frac{(1+\alpha\alpha_0)^2}{2(1+\alpha^2)(1+\alpha_0^2)}<\frac{1}{2}.
\end{gather}
Without the presence of the flat band the condition in Eq.~\eqref{condition} is never satisfied and no DQPT would occur.
Another observation here is that when $\alpha=\alpha_0$ the conditions for a DQPT are not met. From this, we conclude that a simple gap opening is not enough, a change in the hopping ratio is also needed. 
Whenever $\alpha\neq\alpha_0$ there can be a momenta with magnitude $q^*$ so that $p_2(q^*)=1/2$ and the condition in Eq.~\eqref{condition} is satisfied in the momentum interval $(0,q^*]$. There is the possibility that $q^*$ is 
infinite or $p_2(q)<1/2$ for every $q$, if that is the case the condition in Eq.~\eqref{condition} is satisfied for every momentum and as we will see, this changes the nature of the 
DQPT. In order to achieve this the hopping ratios must satisfy:
\begin{gather}
\frac{1+\alpha\alpha_0}{\sqrt{(1+\alpha^2)(1+\alpha_0^2)})} \leq \frac{1}{\delta_S},
\label{plat}
\end{gather}
where $\delta_S=1+\sqrt{2}$ is the silver ratio. The inequality in Eq.~\eqref{plat} originates from the condition $p_2(\infty)\leq 1/2$, and is symmetric 
to the interchange of $\alpha\leftrightarrow\alpha_0$, raising or lowering the hopping ratio during the quench will result in the same DQPT. For example, if $\alpha_0=0$ we must 
raise $\alpha$ above $\alpha_c=\sqrt{2+2\sqrt{2}}\approx 2.2$ and there exists a finite $\alpha$ that satisfy inequality~\eqref{plat} whenever $\alpha_0<1/\sqrt{\delta_S^2-1}\approx 0.45$.
\begin{figure}[t]
\includegraphics[width=4.2cm]{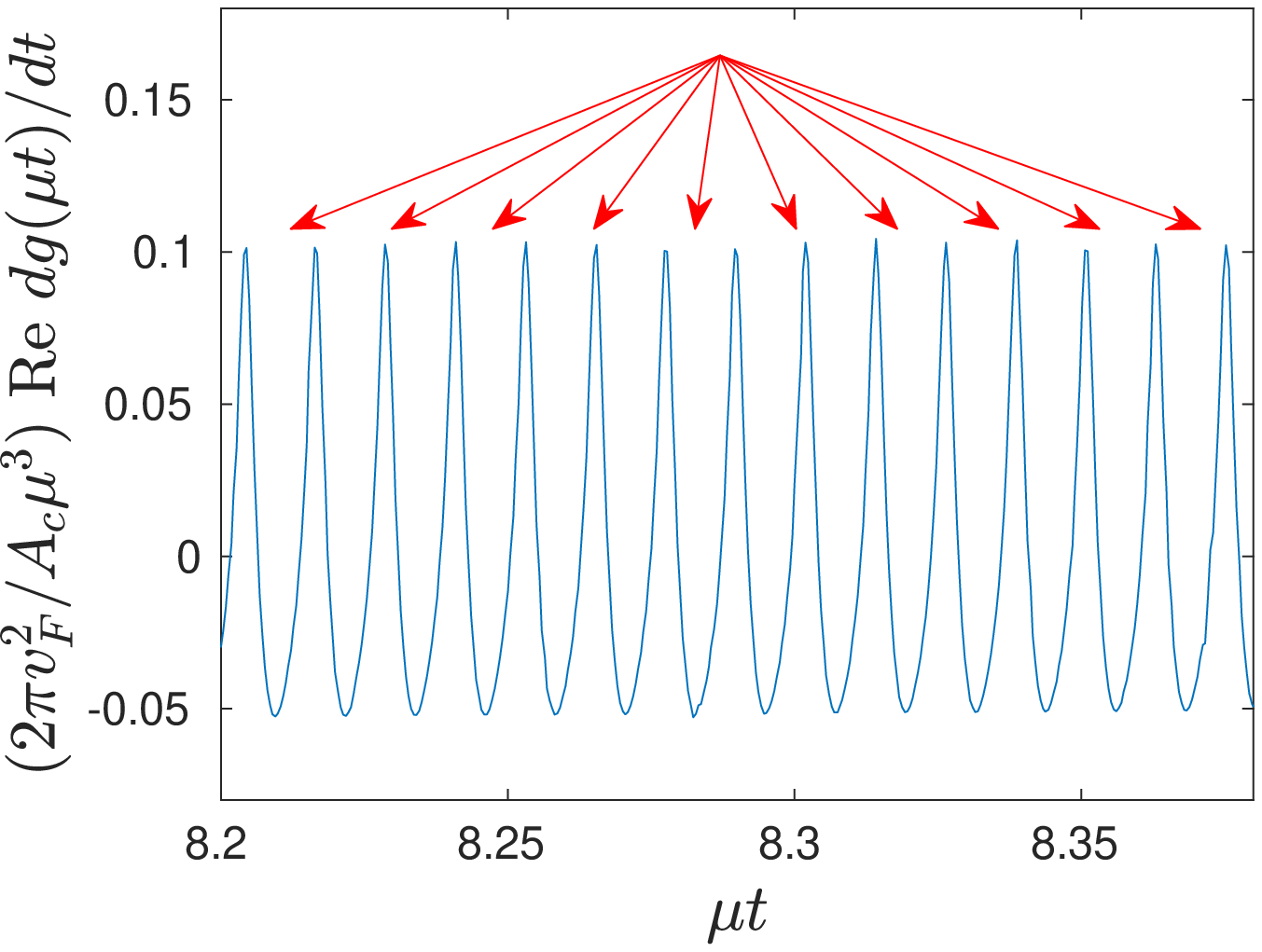}\hspace{0.2cm}\includegraphics[width=4.2cm]{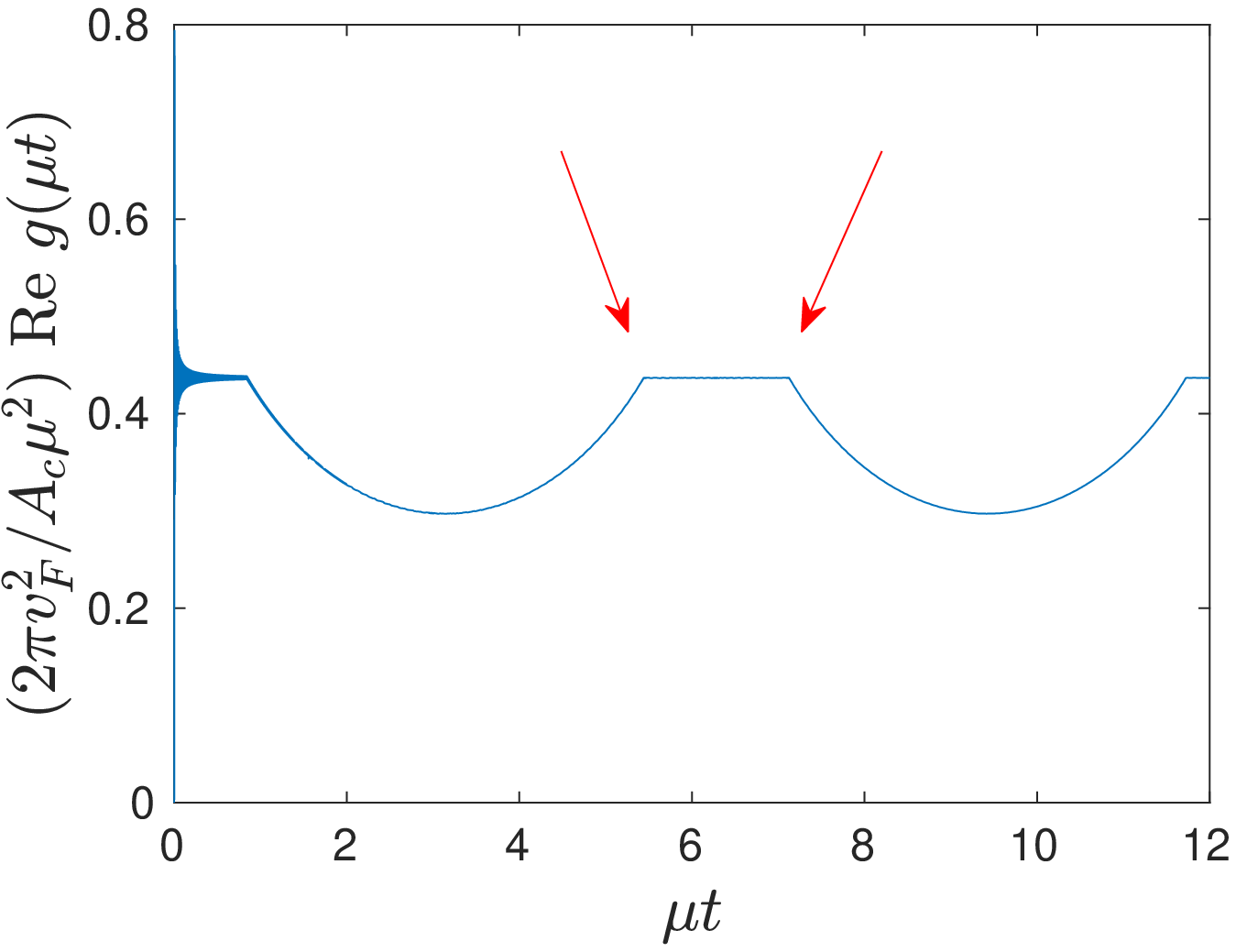}
\caption{The plots of the real part of the universal rate functions for quenches that open a gap $(\mu_0=0\to\mu\neq0)$. Left panel: the first derivative of the rate function when the quench parameters are chosen so that they do not satisfy inequality~\eqref{plat}: $\alpha_0=0.1\to\alpha=1.8$. 
Right panel: the rate function with parameter, so that they satisfy inequality~\eqref{plat}: $\alpha_0=0.1\to\alpha=5$. On both plots the red arrows indicate the kinks that results from the nonanalytic behaviour.}
\label{rateplot}
\end{figure}

When Eq. \eqref{plat} is satisfied, the rate function itself exhibits nonanalytic behaviour, which is visible as kinks on the right hand side of Fig.~\ref{rateplot}. 
The rate function develops plateaus, whose value coincides with the fidelity\cite{fidelity}, namely the overlap between the  groundstates of the pre- and postquench Hamiltonians. 
Its value is calculated from Eq.~\eqref{kisg} by replacing $G(q,t)$ with $p_2$ only. This means that during times corresponding to the plateau, the dominant mode is $p_2$, therefore electrons prefer to reside in the ground state 
of the post quench Hamiltonian. At times between the plateaus all the bands contribute to the rate function with nondominant weights and hence a DQPT occurs.

Whenever Eq. \eqref{plat} is not satisfied the rate function quickly starts to oscillate around the fidelity and approaches it fast with increasing time. 
The DQPTs in this case only occur in the first derivative of the rate function as kinks which are visible in the left panel of Fig.~\ref{rateplot}.
Previously, in two dimensional systems the nonanalyticies were observed only in the first derivative of the rate function\cite{Vajna2015,Heyl2018}. Here, we demonstrated 
that the DQPTs can in fact be observable in the rate function in 2D. 

In either case the prequench state is nontopological, i.e. its Berry curvature integral is zero. After the quench the available bands can have nontrivial topology.
In both cases DQPTs do occur, however the nature of these DQPTs change for values of the parameter $\alpha$ according to \eqref{plat}, which has no connection to the topology of the bands. 
Why does this change happen? 
The simplest explanation is that, since the overlaps can be interpreted as occupation probabilities of the postquench bands, a population inversion 
occurs for the whole system. When there exists a $q^*$ such that $p_2(q^*)=1/2$ only those states can suffer population inversion whose momentum is in the interval $(0,q^*]$. If there is no such $q^*$, that solves the equation $p_2(q^*)=1/2$ or it is infinite then every state of the system partakes in the population inversion. This leads to time intervals when every band can participate in the dynamics with nondominant weights and thus DQPTs appear in the rate function itself.

We remark that the one particle return amplitude depends on the chemical potential in the following way: $G(q,t,\mu)=G(q/\mu,\mu t)$. Thus the rate function scales with $\mu$ as $g(\mu,t)=\mu^2f(\mu t)$, where $f(x)$ is an universal scaling function, for the real part of $f$ see Fig~\ref{rateplot}. or the top figures of Fig.~\ref{X}.
\subsection{Pancharatnam geometric phase and possible dynamical topological order parameter}
\begin{figure}[h]
\includegraphics[width=9cm]{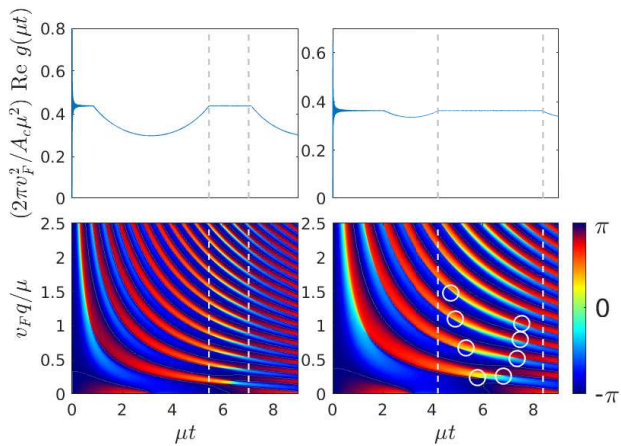}
\caption{The real part of the universal rate function and the Pancharatnam geometrical phase for quenches that open a gap $(\mu_0=0\to\mu\neq0)$. Top row: the plot of the rate functions with parameters, top left: $\alpha_0=0.1\to\alpha=5$ and top right: $\alpha_0=0.1\to\alpha=3.2$. Bottom row: the contourplots of the Pancharatnam geometric phases. 
The dashed lines indicate the exact moment of the DQPTs (gray dashed lines on the top row, white dashed lines on the bottom row), we can observe that the vortices are created pairwise and with different circulation during 
the time interval of the plateaus. With fixed $\alpha_0$ and decreasing $\alpha$ the intervals become larger and they inevitably merge together for parameters that does not satisfy \eqref{plat}. 
For better visibility some vortices are highlighted with white circles in the bottom right panel. }
\label{X}
\end{figure}

The topological nature of DQPTs has been theorized\cite{DTOP} by identifing a dynamical topological order parameter (DTOP), which 
has been connected to the Pancharatnam geometric phase of the return amplitude\cite{DTOP,DTOP2}. In a recent experimental observation of DQPTs\cite{Exp2018}, 
the DTOP were identified by the number of dynamical vortices that were created and annihilated in the Brillioun zone at times of the DQPT. Recasting Eq.~\eqref{loch1} into the following form:
\begin{gather}
G(\mathbf q,t)=R_{\mathbf q}(t)e^{i\varphi_\mathbf q(t)},
\end{gather}
the Pancharatnam geometric phase is available for study. According to rotational symmetry of the system at low energies, 
this phase only depends on the magnitude of the momentum: $\varphi_\mathbf q(t)=\varphi(q,t)$. 
The contourplot of Fig.~\ref{X}. reveals that in the time-momentum space $(t,q)$ vortices are created at finite times whenever the quench parameters obey inequality~\eqref{plat}. We argue that the number of these vortices 
can be classified as a DTOP. As time passes and reaches the exact moment of a kink before a plateau in the rate function, a countable infinitely many vortices are created. 
For small momenta the vortices reside close to the middle of the time interval of the plateau. With increasing momentum, however, the vortices appear and disappear closer to the beginning and ending time of the plateau. 
The exact moment for the DQPT to appear is an accumulation point of the vortices in the momentum-time plane. Every vortex with positive circulation adds $+1$ to the DTOP, hence it becomes nonzero. 
By the time we are at the end of the plateau these vortices move to a different momenta and their circulation is shifted to $-1$. Thus at the exact moment of the cusp at the end of the plateau the sum of the number of vortices with respect to circulation becomes zero again.

As stated above 
whenever $\alpha_0<1/\sqrt{\delta_S^2-1}$ there exists a finite $\alpha$ for which inequality~\eqref{plat} holds. With decreasing $\alpha$, it inevitably reaches the point in the $\alpha_0-\alpha$ parameter space (Fig.~\ref{paraspace}) 
when \eqref{plat} is not satisfied, and the vortices at the times of the different kinks merge together. We can interpret this with the argument that in this case the vortices are created at the infinite far points of $q\to\infty$ and they disappear 
at the infinite far points of $t\to\infty$. The system has at all times a nonzero unchanged DTOP and the rate function itself does not develop kinks and the DQPTs are only present in the first derivative.

\subsection{Macroscopically different dynamical phases separated by the DQPTs}
During a conventional equilibrium phase transition\cite{sachdev}, the nonanalytic behaviour of the free energy density separates phases of the system, that have different macroscopical properties. 
The theory of DQPTs rely on the existence of nonanalyticies in the dynamical free energy as a function of time, however these are usually not accompanied by changes in the macroscopical properties.

Within the low energy discription of the present model, we can argue that the nonanalytic kinks in the rate function separates macroscopically different phase structures in the momentum-time plane. 
As stated above, during the time interval of a plateau, the vortices appear and disappear with the exact moment of the DQPT serving as an accumulation point for the positions of these vortices. The number of
vortices is proportional to $\alpha v_Fq_F/\mu$ for finite $q_F$. 
As the cutoff is taken to infinity  the developed ladder structure becomes infinitly large and in this sense it can be considered macroscopical.
At the moment of DQPT this macroscopic structure appears or disappears.
 Hence the nonanalytic kink in the rate 
function heralds a macroscopical change in one of the properties of the system, making it a true dynamical phase transition. 


\subsection{Robustness of DQPTs in the rate function}

The nonanalytic kinks that appear in the rate function follow directly from the continuous rotational symmetry of the system, valid at low energies.
In this limit, the system behaves as an effective one dimensional system, explaining the kinks in the rate function itself.
Moving away from the linearized Hamiltonian in Eq.~\eqref{ham} and including higher order terms of the momentum from the tight binding dispersion would 
lower the continuous rotational symmetry to discrete 3-fold rotational symmetry. In this case the vortices of Fig.~\ref{X} 
in the momentum-time space would distort and the vortex dynamics would become anisotropic in the $q_x-q_y$ plane as well. This would eventually lead the 
cusps at the beginning and the end of the plateaus to be rounded. The nonanalytic kinks would appear only in the first 
derivative of the rate function at times, when the rounding of the cusp begins and ends. Due to the lowering of the symmetry 
the structure of the Fisher zeroes would also change from being on a line to being on an area in the complex plane. However, 
if we stay within the realm of the low energy effective theory, such that $\mu$ is much smaller than the bandwidth, the rounding would be very small and any anisotropy in the  vortex dynamics would be hardly observable.
This is analogous to the effect of trigonal warping on the Fermi surface in graphene and its effect on physical observables such as the optical conductivity\cite{trigonal2,trigonal1}.

\section{Quench within the gapped phase}\label{gap1}
\begin{figure}[h]
\includegraphics[width=4.2cm]{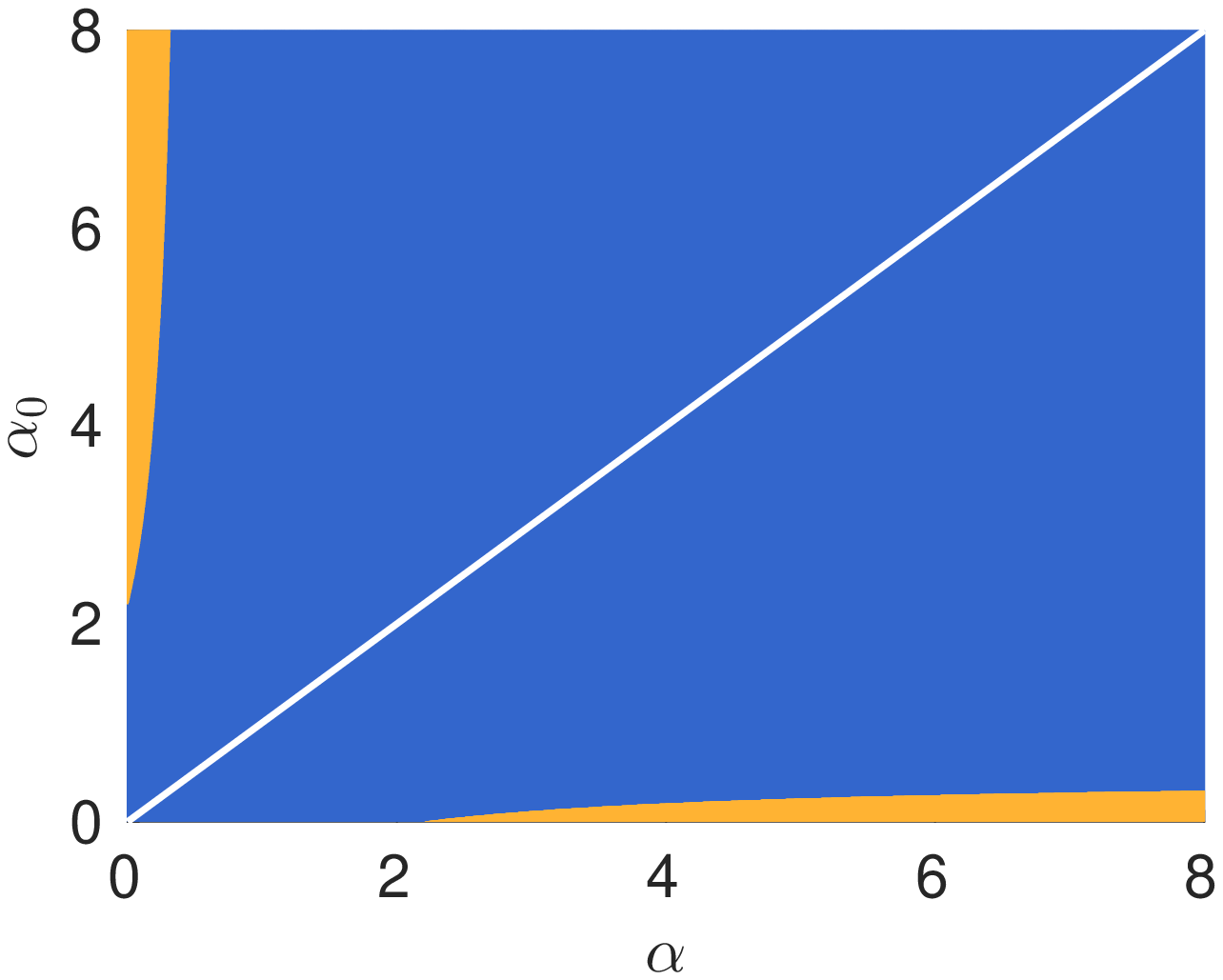}\hspace{0.2cm}\includegraphics[width=4.2cm]{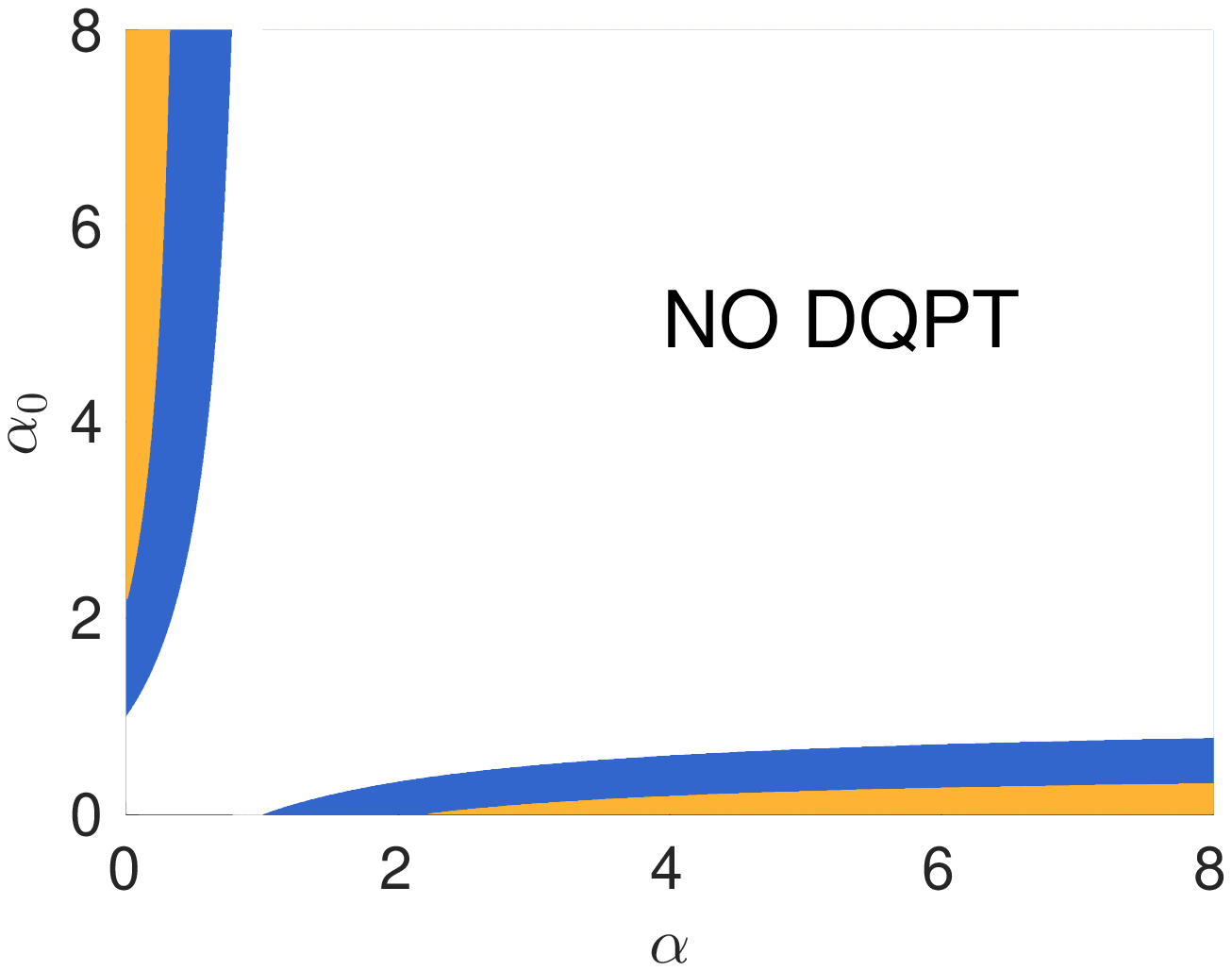}
\caption{The $\alpha_0-\alpha$ parameter space, where the colors depict the 
different regions whether DQPTs cannot occur at all (white), DQPTs occur in the 
first derivative of the rate function only (blue) and DQPTs occur in the rate function itself (orange). The left panel shows the regions during a gapless to gapped quench scenario $\mu_0=0\to\mu\neq0$ and the right panel corresponds to quenches within the same 
gapped phase $\mu_0=\mu$.}
\label{paraspace}
\end{figure}

The next scenario we consider is when there is a finite gap in the prequench Hamiltonian ($\mu_0\neq0$) and it is not closed during the quench ($\mu\neq0$). 
As was the case before phase ergodicity holds and we only need to inspect  the overlaps:
\begin{gather}
p_0=\frac{(\alpha_0-\alpha)^2v_F^2q^2}{(1+\alpha^2)(E_0^2+v_F^2(1+\alpha_0^2)q^2)},\nonumber\\
p_{1,2}=\frac{(E_{1,2}E_0+v_F^2q^2(1+\alpha\alpha_0))^2}{(E_0^2+v_F^2(1+\alpha_0^2)q^2)(E_{1,2}^2+v_F^2(1+\alpha^2)q^2)},
\label{m0nem0}
\end{gather}
where $E_0=\mu_0/2-\sqrt{\mu_0^2/4+v_F^2(1+\alpha_0^2)q^2}$. 
The flat band overlap decreases with the momentum, $\partial p_0/\partial q\leq0$ and $p_0(0)=2p_0(\infty)$ with $p_0(\infty)<1/2$. The high energy band overlap $p_1$ has two zeroes: $p_1(0)=p_1(q')=0$ and increases with momentum 
on the $[q',\infty)$ domain: $\partial p_1/\partial q>0$, if $q>q'$. Due to $p_1(0)=0$, we immediately notice that if $\alpha$ and $\alpha_0$ are close $p_2$ will always be above $1/2$, thus the conditions for DQPTs are not met. 
In the presence of a gap, the prequench state is topological and by changing the hopping ratio the available bands all have different integrals of the Berry curvature. We conclude that no DQPTs occur in spite that the quench reaches a different topological phase. 
However, there are certain $\alpha_0,\alpha$ parameters for which nonanalytices can occur in the first derivative of the rate function, as well as in the rate function itself. This is shown in Fig.~\ref{paraspace}.
In order to observe DQPTs in the rate function, the condition is $p_2(\infty)\leq 1/2$, which translates into the same inequality~\eqref{plat} for the hopping ratios as before. For hopping ratios $\alpha,\alpha_0$ that satisfy inequality \eqref{plat} 
the low energy band overlap of the pre- and postquench Hamiltonian $(p_2)$ is always below $1/2$, and population inversion occurs for all momentum. This time however, there exists a momentum $q^*$, with $p_0(q^*)=1/2$ and the condition \eqref{condition} 
is satisfied for the interval $[q^*,\infty)$. 
Similarly the Pancharatnam geometric phase in the time-momentum space also develops pairwise vorteces with different circulation giving rise to a DTOP. These vortices only appear for momentum larger than $q^*$.

\section{Closing the gap}\label{gap2}

Finally case we discuss the gapped to gapless quench, i.e. the closing of the gap during the quench: $\mu_0\neq0\to\mu=0$. 
In this scenario the postquench bands are rationally dependent: $E_0+E_1+E_2=0$. The return amplitude is the product with respect to the momentum $q$ of the quantity:
\begin{gather}
G(q,t)=p_0+(p_1+p_2)\cos E_1t+i(p_2-p_1)\sin E_1t.
\label{Gqt}
\end{gather}
The condition for DQPTs is $G(q,t^*)=0$, for $q\neq0$. In this case $p_2\neq p_1$, $\forall q\neq0$ and $t^*$ can be exactly calculated:
\begin{gather}
t_n^*=\frac{(2n+1)\pi}{E_1(q^*)},
\end{gather}
where $q^*$ is the momentum when $p_0(q^*)=1/2$ and $p_0$ is given in Eq.~\eqref{m0nem0}. With some algebra we determine $q^*$ to be:
\begin{gather}
q^*=\frac{\mu_0}{2v_F\sqrt{1+\alpha_0^2}}\sqrt{\frac{(\alpha-\alpha_0)^4}{(1+\alpha\alpha_0)^4}-1}.
\end{gather}
This momentum has to be real, thus in order to observe a DQPT the hopping ratios must satisfy the following inequality:
\begin{gather}
|\alpha-\alpha_0|-\alpha\alpha_0>1.
\label{condi}
\end{gather}
We note that  all postquench bands are nontopological and the prequench state has zero \jav{Berry curvature integral} only when $\alpha_0=1$. According to \eqref{condi} DQPTs cannot occur when $\alpha_0=1$, thus in 
this case changing the topology is a necessary but not sufficient condition.

\section{Conclusion} 
The $\alpha-T_3$ lattice model can be realized experimentally with cold fermionic atoms loaded into an optical lattice. 
Following the proposal for the optical dice lattice ($\alpha=1$)\cite{Rizzi2006}, one simply needs to dephase one of the three pairs of laserbeams to obtain $\alpha\neq1$\cite{Raoux2014}. 
Of late the detection of DQPTs have seen much success with different experimental platforms such as quantum simulators and nanomechanical systems\cite{Jiangfeng2019,Exp2018,Jurcevic2017}. 
Furthermore, the recent experimental success of observing DQPTs using single photon resonators\cite{Xue2019} can also allow the possibility to simulate the different quench scenarios in the Hamilton matrix of Eq.~\eqref{ham}.

We investigated different geometrical quench scenarios for the $\alpha-T_3$ lattice model and studied the conditions for the occurance of dynamical quantum phase transitions. 
Either using the fully filled low energy band or the fully filled low energy and flat band as the prequench state, yield the same identical results.
\jav{The presented model has rich topological properties, yet the appearance of DQPTs are not tied to them for quenches that open a gap or quenches within the gapped phase. 
We showed that the basic condition for DQPT to happen is population inversion. Since momentum is a good quantum number, an initial 
 occupied state labeled by the momentum $\mathbf q$ exhibits DQPT when it undergoes population inversion. 
If population inversions occur for a finite momentum interval, the DQPTs are only present in the derivative of the rate function. 
However, if the momentum interval, for which Eq.~\eqref{condition} is fulfilled, is infinite the DQPTs also appear in the rate function itself.
The Pancharatnam geometric phase of the rate function in the time-momentum space also features a qualitative change of properties whenever 
the momentum interval for population inversions is infinite. 
With inequality Eq. \eqref{plat} satisfied, vortices form pairwise and with different circulation. 
Interestingly, the positions of the vortices form an infinite vortex ladder, that can be interpreted 
as a macroscopic structure in the momentum-time plane that appears and disappears at a DQPT. For parameters that does not satisfy Eq. \eqref{plat} these 
vortices are absent at finite times. Finally, closing the gap during a quench can also support DQPTs, however this time the DQPT happen for the single momenta where the population inversion occurs, i.e. $p_0(q^*)=p_1(q^*)+p_2(q^*)$. Coincidentally, in this case the change of topological properties during the quench is a necessary condition. }

\begin{acknowledgments}
This research is supported by the National Research, Development and Innovation Office - NKFIH within the Quantum Technology National Excellence Program (Project No.
      2017-1.2.1-NKP-2017-00001), K119442, by the  BME-Nanonotechnology FIKP grant of EMMI (BME FIKP-NAT).

\end{acknowledgments}

\flushend
\bibliographystyle{apsrev}
\bibliography{refgraph}

\end{document}